\documentclass[reqno,a4paper]{article}
\usepackage{amsmath,amssymb,amsfonts,a4,cite}
\numberwithin{equation}{section}

\def\Mult#1#2#3{{\genfrac{[}{]}{0pt}{0}{#1}{#2}}_{#3}}

\def\h{\theta}
\def\ra{\rho_1}
\def\r2{\rho_2}
\def\({\biggl(}
\def\){\biggr)}
\def\b#1{\bar{#1}}

\begin{document}
\title{Central Charge and the Andrews-Bailey Construction}

\author{Leung Chim
\thanks{
e-mail: chim@mundoe.maths.mu.oz.au}\\
{}\\
{\it Department of Mathematics,}\\
{\it The University of Melbourne,}\\
{\it Parkville, Victoria, 3052,}\\
{\it Australia.}\\
{}\\
{Research Report No.7}}

\maketitle

\begin{abstract}
From the equivalence of the bosonic and fermionic representations
of finitized characters in conformal field theory, one can extract
mathematical objects known as Bailey pairs. Recently Berkovich, McCoy
and Schilling have constructed
a `generalized' character
formula depending on two parameters $\ra$ and $\r2$, using the Bailey
pairs of the unitary model $M(p-1,p)$.
By taking appropriate limits of these parameters, they were able to
obtain the characters of model $M(p,p+1)$, $N=1$
model $SM(p,p+2)$, and the unitary $N=2$ model with central charge
$c=3(1-{\frac{2}{p}})$. In this letter we computed the
effective central
charge associated
with this `generalized' character formula using a saddle point
method. The result is a simple expression in dilogarithms which
interpolates between the central charges of these unitary models.
\end{abstract}

\section{Introduction}

More than a decade since its creation, two dimensional
conformal field theory (CFT) \cite{1}
and its integrable perturbations \cite{2} still remain as one of
the most active research topics in modern physics. A current
focus is in the study of various bases of the Hilbert space in CFT.
Different choices of the basis would lead to a different representation
for the partition function of the CFT defined on a compact
manifold such as a torus or a cylinder.
This partition function is usually written
in terms of characters of the Virasoro or some extended
algebras. The `bosonic' form of the character formula is well known for
quite some time \cite{3}. Recently the Stony Brook group have constructed
numerous new character formulae based on fermionic quasi-particles \cite{4,5,BMS}.
For several CFTs, more than one `fermionic' expression
exist for the same conformal character. In these cases,
the different expressions are related to the different integrable
perturbations of the same CFT. These developments all lend supports to the idea of a
massless scattering
S-matrix description of CFT \cite{6,7,8,13}.
The construction of the quasi-particle
basis of the Hilbert space is also apparently related to the problem of
diagonalizing the infinite set of local Integrals of Motion in CFT
\cite{9,FRS}.
For a description in terms of other bases see \cite{10,11,12}.

The equivalence of the bosonic and fermionic character formulae give rise
to beautiful $q$-series identities of the Rogers-Ramanujan type \cite{15}. To
prove
some of these identities,
the method of finitization of character formula was
employed in \cite{16}. The basic idea is that the equivalence between bosonic
and fermionic finitized characters would also imply the equivalence of the
$q$-series \cite{schur}.
In \cite{17}, several classes of $q$-series identities were proven using
Andrews' generalization \cite{18} of Bailey's lemma \cite{19}.
The key observations in \cite{17} was that
Bailey pairs can be extracted from finitized characters%
\footnote{To extract Bailey pairs, the finitized characters must have the
general bosonic form of \cite{FB}. We would to thank B. McCoy for this
comment.}%
(such as those of \cite{16}), and several 
series of CFT
are `linked' on a so-called Bailey's chain \cite{18}.
The equivalence proof for all members of a series is a
straight
forward application of Bailey's lemma, once a proof is established for a
single
member \cite{18}.

In a remarkable paper \cite{20},
the procedure of \cite{17} was repeated using a more
general form of the Andrews-Bailey construction which contains
two parameters $\ra$ and $\r2$ \cite{18}.
From the (dual) Bailey pairs for the
unitary minimal model $M(p-1,p)$,
the more general construction gives a `generalized' character formula
(equation (4.19) of \cite{20})
depending
on the two parameters.
Three specializations of these parameters lead to
known results:
\begin{equation}
\label{e:case}
\begin{align}
\text{(I):}&\: \ra \rightarrow \infty, \quad \r2 \rightarrow \infty; \notag\\
\text{(II):}&\: \ra \rightarrow \infty, \quad \r2 = finite;\\
\text{(III):}&\: \ra = finite, \quad \r2 = finite. \notag
\end{align}
\end{equation}
In the first case, the `generalized' character formula becomes the
character for the next model in the unitary series, ie $M(p,p+1)$
with central charge $1-\frac{6}{p(p+1)}$.
Case (II) leads to fermionic character formula for $N=1$ supersymmetric
model
$SM(p,p+2)$ with central charge
$\frac{3}{2}-\frac{12}{p(p+1)}$,
while case (III) gives the fermionic character of unitary
$N=2$
model with central charge $c=3(1-{\frac{2}{p}})$.
It is amazing that this `generalized' character formula
connects the unitary models with
their supersymmetric counterparts. In fact, this construction can
also be applied to any minimal models $M(p,p^{\prime})$ \cite{17,BMS,22}.
A natural question (raised in \cite{20}) which needs addressing is whether
this construction has any connection to massless renormalization
group flows between these CFT \cite{6, 23}.

In this letter, we shall attempt to understand this `generalized'
character formula for the unitary series
by computing the associated `generalized'
effective central charge.
In \S 2 we give a brief
review of the construction detailed in \cite{20} and establish our notations.
The `generalized' effective central charge is calculated in \S 3 via a
saddle point approximation following \cite{24,25,4}. A disscusion
of our result is given in \S 4 where we speculate on a
possible interpretation of the `generalized' character formula.

\section{The Andrews-Bailey construction of the unitary models}

The unitary CFT $M(p-1,p)$ is the continuum limit of the $(p-1)$-states
RSOS lattice model \cite{26} at its critical point between
Regime III and IV \cite{27}.
The equivalence of the associated bosonic and fermionic finitized
characters can be written as \cite{16}
\begin{equation}\label{e:BF}
B_{r,s}^{(L,p)} = F_{r,s}^{(L,p)}.
\end{equation}
The bosonic side has the form
\begin{equation}\label{e:bc}\begin{split}
B_{r,s}^{(L,p)}(q) 
= \sum^{\infty}_{j=-\infty}
\(&q^{j(jp(p-1)+pr-(p-1)s)}\Mult{L}{[{\frac{1}{2}}
(L+s-r)]-pj}{q}\\
&-q^{(jp-s)(j(p-1)-r)}\Mult{L}{[{\frac{1}{2}}
(L-s-r)]+pj}{q} \),\\
\end{split}\end{equation}
where $[n]$ denotes the integer part of $n$, and
\begin{equation}
\Mult{n}{m}{q}=
\begin{cases}
{\frac{(q)_n}{(q)_m{(q)_{n-m}}}}& \text{for $0 \le m \le n$}\\
0& \text{otherwise},
\end{cases}
\end{equation}
is the usual $q$-binomial coefficient with
\begin{equation}
(a)_n = \frac{(a)_{\infty}}{(aq^n)_{\infty}};
\quad (a)_{\infty} = \prod_{l=0}^{\infty}(1-aq^l).
\end{equation}
One should note that this finitized character is equal to 
the off-critical corner
transfer matrix of the RSOS model in its
low-temperature Regime III, defined on a square
lattice of size $L$ \cite{26}.
In the limit $L \rightarrow \infty$,
\eqref{e:bc} becomes the character formula for the irreducible
representation generated
by the primary field $\Phi_{(r,s)}$ \cite{3}, with normalization
$\chi_{r,s}(q)=1+\sum_{N\ge1}a_{N}q^N$.

There are two forms for the fermionic character $F_{r,s}^{(L,p)}$, depending
on the finite parameter $L$. Let $C_{p-3}$ and $I_{p-3}$ stand for respectively
the Cartan
and incidence matrices of the Lie algebra $A_{p-3}$. Furthermore
denote the $i$ unit vector in ${\mathbb{R}}^{\,p-3}$ as $\vec{e}_i$, and
set $\vec{e}_i = \vec{0}$ for $i<1$ or $i>(p-3)$. 
Then
\begin{equation}
\begin{split}
F_{r,s}^{(L,p)}(q) = q^{-\frac{1}{4}(s-r)(s-r-1)}
&\sum_{\vec{m}\in2{\mathbb{Z}}^{p-3}+\vec{Q}_{r,s}}
q^{\frac{1}{4}\vec{m}^T{C_{p-3}}\vec{m}-\frac{1}{2}\vec{A}_{r,s}\vec{m}}\\
&\times\prod_{i=1}^{p-3}\Mult{\frac{1}{2}(I_{p-3}\vec{m}+\vec{u}_{r,s}
+L\vec{e}_1)_i}{m_i}{q},\\
\end{split}
\end{equation}
where $\vec{m}^T=(m_1,\dots,m_{p-3})$.
When $L+r-s$ is even
\begin{align}
\vec{A}_{r,s}& =\vec{e}_{s-1} \notag\\
\vec{u}_{r,s}& =\vec{e}_{s-1} + \vec{e}_{p-r-1}\\
\vec{Q}_{r,s}& =(r-1){\sum_{i=1}^{p-3}\vec{e}_i} +
(\vec{e}_{s-2}+\vec{e}_{s-4}+\dots) +
(\vec{e}_{p-r}+\vec{e}_{p+2-r}+\dots); \notag
\end{align}
and when $L+r-s$ is odd,
\begin{align}
\vec{A}_{r,s}& =\vec{e}_{p-s-1} \notag\\
\vec{u}_{r,s}& =\vec{e}_{p-s-1} + \vec{e}_{r}\\
\vec{Q}_{r,s}& =(s-1){\sum_{i=1}^{p-3}\vec{e}_i} +
(\vec{e}_{r-1}+\vec{e}_{r-3}+\dots) +
(\vec{e}_{p-s}+\vec{e}_{p+2-s}+\dots). \notag
\end{align}
These two forms yield the same $q$-series in the limit $L\rightarrow\infty$.
Proofs of the fermionic sums are given in \cite{Berkovich,28,ole}.

Two sequences $\{\alpha_n\}$ and $\{\beta_n\}$ form a (bilateral) Bailey pair relative
to $a$ if they satisfy the relation 
\begin{equation}\label{e:blemma}
\beta_n = \sum_{j=-\infty}^{n}\frac{\alpha_j}
{(q)_{n-j}(aq)_{n+j}}.
\end{equation}
If we set $L=2l+r-s+2x$, then from \eqref{e:BF} we can read off a (bilateral)
Bailey pair relative to $a=q^{r-s+2x}$ as
\begin{subequations}
\label{e:bpair}
\begin{align}
\alpha_n& =\begin{cases}
	q^{j(jp(p-1)+pr-(p-1)s)}& \text{for $n=pj-x$} \\
	-q^{(jp-s)(j(p-1)-r)}& \text{for $n=pj-r-x$} \\
	0& \text{otherwise}
	\end{cases} \\
\beta_n& =\begin{cases}
	\frac{1}{(aq)_{2n}}F_{r,s}^{(2n+r-s+2x,p)}(q)& \text{for $n\ge0$}\\
	0& \text{otherwise.}
	\end{cases}
\end{align}
\end{subequations}

An important step in the Andrews-Bailey construction of the unitary models
is to define another Bailey pair relative to $a$
\begin{subequations}
\label{e:dbpair}
\begin{align}
A_n& =\begin{cases}
	q^{j^2p-sj+x(s-r-x)}& \text{for $n=pj-x$} \\
	-q^{j^2p-sj+x(s-r-x)}& \text{for $n=pj-r-x$} \\
	0& \text{otherwise}
	\end{cases} \\
B_n& =\begin{cases}
	\frac{1}{(aq)_{2n}}q^{n^2}a^{n}F_{r,s}^{(2n+r-s+2x,p)}(q^{-1})&
	\text{for $n\ge0$}\\
	0& \text{otherwise},
	\end{cases}
\end{align}
\end{subequations}
which are dual to \eqref{e:bpair} \cite{18,20}.
Here
\begin{equation}
\label{e:dualF}
\begin{split}
F_{r,s}^{(L,p)}(q^{-1}) = q^{\frac{1}{4}(s-r)(s-r-1)}
&\sum_{\vec{m}\in2{\mathbb{Z}}^{p-3}+\vec{Q}_{r,s}}
q^{\frac{1}{4}\vec{m}^T{C_{p-3}}\vec{m}+\frac{1}{2}(\vec{A}_{r,s}
-\vec{u}_{r,s}-L\vec{e}_1)\vec{m}}\\
&\times\prod_{i=1}^{p-3}\Mult{\frac{1}{2}(I_{p-3}\vec{m}+\vec{u}_{r,s}
+L\vec{e}_1)_i}{m_i}{q},
\end{split}
\end{equation}
This dual transformation $(q\rightarrow{q^{-1}})$
take us from the $M(p-1,p)$ finitized characters to
the $M(1,p)$ finitized characters \cite{5,ole}. The non-unitary
minimal model $M(1,p)$ has actually zero operator content in the usual range
of $r$ and $s$%
\footnote{Although one can extend these ranges to generate interesting CFTs
\cite{29,30}.}%
, but admits nontrivial finitizations%
\footnote{In fact, up to some prefactors, \eqref{e:dualF} is 
the finitization of the ${\mathbb{Z}}_{p-2}$ parafermions
which describe the critical point between Regime I and II of the RSOS
model \cite{26,31}. We would like to thank O. Foda and
O. Warnaar for pointing this out.}%
.

The Andrews-Bailey construction tells us that if \eqref{e:dbpair}
is a (bilateral) Bailey pair, then
\begin{subequations}
\begin{align}
A_{n}^{\prime}& =\({\frac{(\ra)_n(\r2)_n(aq/\ra\r2)^n}
		{(aq/\ra)_n(aq/\r2)_n}}\)A_n\\
B_{n}^{\prime}& =\sum_{m=-\infty}^{n}
\({\frac{(\ra)_m(\r2)_m(aq/\ra\r2)_{n-m}(aq/\ra\r2)^m}
{(q)_{n-m}(aq/\ra)_n(aq/\r2)_n}}\)B_{m}
\end{align}
\end{subequations}
also forms a (bilateral) Bailey pair with respect to $a$. 
Now using the defining relation \eqref{e:blemma} with this new Bailey
pair and taking the limit $n\rightarrow\infty$, one easily obtains
the formula
\begin{multline}
\label{e:genchar}
\frac{(aq/\ra)_{\infty}(aq/\r2)_{\infty}}{(aq)_{\infty}(aq/\ra\r2)_{\infty}}
\sum_{j=-\infty}^{\infty}q^{j(jp-s)+x(s-r-x)}\\
\times\({\frac{(\ra)_{pj-x}(\r2)_{pj-x}(aq/\ra\r2)^{pj-x}}
{(aq/\ra)_{pj-x}(aq/\r2)_{pj-x}}
-\frac{(\ra)_{pj-r-x}(\r2)_{pj-r-x}(aq/\ra\r2)^{pj-r-x}}
{(aq/\ra)_{pj-r-x}(aq/\r2)_{pj-r-x}}}\)\\
=\sum_{n=0}^{\infty}(\ra)_n(\r2)_n(aq/\ra\r2)^n
\frac{q^{n^2}a^n}{(aq)_{2n}}F^{(2n+r-s+2x,p)}_{r,s}(q^{-1}).
\end{multline}
We shall refer to the expression in \eqref{e:genchar} as the `generalized'
character formula, and we will compute the effective central charge
associated with it in the next section.
In the limiting case (I) (and setting $x=0$),
\eqref{e:genchar} becomes the character formula
for $\chi_{s,r}^{(p,p+1)}$, where $1\le r \le(p-2)$ and $1\le s \le(p-1)$.
Similarly, the `generalized' character \eqref{e:genchar} yields characters
for the $N=1$ supersymmetric model $SM(p,p+2)$ in the case (II), while
case (III) leads to characters of the $N=2$ model with central charge
$c=3(1-\frac{2}{p})$ \cite{20}. Repeating the Andrews-Bailey
construction starting from \eqref{e:BF} and taking $L=2l+r-s+2x+1$, another
`generalized' character can be obtained. The latter becomes the character
formula for $\chi_{s,r+2}^{(p,p+1)}$ in case (I), and gives more characters
for the supersymmetric models in the cases of (II) and (III) (please see
\cite{20} for more details). Since this second `generalized' character
leads to the same central charge as \eqref{e:genchar}, it will not be considered
further in this work.

\section{Effective central charge}

In this section, we shall calculate the asymptotic behavior of
\eqref{e:genchar} as $q\rightarrow1^{-}$.
This method of computing the effective central charge for CFT
fermionic characters are by now standard \cite{24,25,4,28}.
Therefore we will be brief with the procedure, but detailing in
places where our calculation differs from the norm.
Firstly we shall predict the asymptotic growth of the `generalized'
character
using a physical argument. Subsequently we will derive this
asymptotic behavior directly from the character formula.

\subsection{Asymptotic behavior of the `generalized' character}

Characters in CFT admit an interpretation as the partition function of
the model defined on a cylinder with
conformal boundary conditions on its
rims \cite{34}. The modular invariance
property of these character formulae
give us precise information about their asymptotic behavior. 
The working assumption in
this section will be that \eqref{e:genchar}
also gives the cylindrical
partition function for some 
quantum field theory with appropriate
boundary conditions labeled $a$ and $b$. 
Noted that $a$ and $b$ depend on the values of $r$ and $s$, as well as
$\ra$ and $\r2$.
Define the modular parameter
\begin{equation}
q=e^{2\pi{i}\tau}; \quad \tilde{q}=e^{-2\pi{i}/\tau};
\quad \text{with} \quad \tau=\frac{iR}{2\pi{L}}
\end{equation}
for a cylinder of length $L$ and circumference $R$.
If we take the (imaginary) time coordinate to be in the $R$
direction and space in the $L$ direction,
the generalized character \eqref{e:genchar} can be written as
\begin{equation}
\chi_{s,r}^{(p)}(\ra,\r2;\tau)
= Tr_{P}e^{-R\,{\mathcal{H}}_{ab}(\ra,\r2|p)/L}
\quad = Tr_{P}\:q^{{\mathcal{H}}_{ab}(\ra,\r2|p)},
\end{equation}
where ${\mathcal{H}}_{ab}$ is the (normalized) dimensionless
Hamiltonian of the field theory
with open boundary conditions $a$ and $b$.
The trace is taken over the sector
of the Hilbert space with boundary
condition $P$ along the circumference of the cylinder.
For the cases \eqref{e:case}, ${\mathcal{H}}_{ab}$ becomes
$L_{0}-\Delta(s,r)$ where $\Delta(s,r)$ is the appropriate
conformal dimension for each unitary model.
If instead we take space to be compactified in the $R$ direction
and time to evolve in the $L$ direction, then the partition
function will be%
\footnote{To be more precise, there should be a prefactor on the right
hand side of \eqref{e:innerproduct} involving powers of $q$
due to normalization. However
it becomes irrelevant to our analysis in the limit $q\rightarrow{1^{-}}$.}%
\begin{equation}
\label{e:innerproduct}
\chi_{s,r}^{(p)}(\ra,\r2;\tau)
= \langle{a}|\,e^{-L\,{\mathcal{H}}_{P}(\ra,\r2|p)/R}\,|{b}\rangle,
\end{equation} 
where ${\mathcal{H}}_{P}$ is the dimensionless Hamiltonian with closed boundary
condition $P$.
Here $\langle{a}|$ and $|{b}\rangle$
represent the boundary
states at the ends of the cylinder.
In the limit $L\rightarrow\infty$, the inner product \eqref{e:innerproduct}
is dominated by the ground state of ${\mathcal{H}}_P$ with energy $E_0$
\begin{equation}\label{e:asym}
\lim_{L\rightarrow\infty}\chi_{s,r}^{(p)}(\ra,\r2;\tau)
\sim g_a\,g_b\,{\tilde{q}}^{E_0/{4{\pi}^2}},
\end{equation}
where we denote the contributions from each boundary as $g_a$ and $g_b$.
\eqref{e:asym} is our prediction of the asymptotic behavior of \eqref{e:genchar}.

The fermionic form of the generalized character
\eqref{e:genchar} is most suitable for taking
the asymptotic limit $q\rightarrow1^{-}$.
The important thing to notice here is that
for fixed $p$, $E_0$ depends only on the
parameters $\ra$ and $\r2$, and is independent of $r$ and $s$.
Standard arguments
\cite{4} (related to the $r$ and $s$ independence of $E_0$) give us
the freedom to remove
the restriction ${\vec{Q}}_{r,s}$ and
linear terms in the exponent of $q$
from the fermionic sum in this limit.
Thus to compute the asymptotic behavior of the generalized character,
i.e. to obtain the leading exponent of $\tilde{q}$,
we could just concentrate on the simplest case of the identity
representation
$\chi_{1,1}^{(p)}(\ra,\r2;\tau)$. To implement
the special limits \eqref{e:case} in this case, let us
parameterizes $\ra$ and $\r2$ as
\begin{equation}
\ra = -\frac{q^{\frac{1}{2}}}{A},
\quad \r2 = -\frac{q^{\frac{1}{2}}}{B},
\end{equation}
thus we have%
\footnote{With our choice of parameterization, the limits
\eqref{e:case} actually become (I):$\ra\rightarrow-\infty$,
$\r2\rightarrow-\infty$ and (II):$\ra\rightarrow-\infty$,
$\r2=finite$. However we can still obtain the same
conformal characters from \eqref{e:genchar}.}%
\begin{subequations}
\label{e:case2}
\begin{align}
\text{(I):}&\: A=0, B=0; \notag\\
\text{(II):}&\: A=0, B=1;\\
\text{(III):}&\: A=1, B=1, \notag
\end{align}
\end{subequations}
with $x=0$ and $a=1$ for all three cases.
The limits (II) and (III) lead to the
Neveu-Schwarz characters for the supersymmetric models.
We will not consider the Ramond sector, although
it can be treated by a straightforward generalization of the
computation presented here.
A convenient parameterization of the ground state energy $E_0$
is
\begin{equation}
\label{e:Ec}
E_0(A,B|p) = -\frac{{\pi}^2}{6}\,\tilde{c}(A,B|p).
\end{equation}
From \eqref{e:asym}, we shall interpret $\tilde{c}$
as the `generalized' effective central charge,
and expect that in the limits \eqref{e:case2},
it will take on the values of $1-\frac{6}{p(p+1)}$,
$\frac{3}{2}-\frac{12}{p(p+1)}$
and $c=3(1-{\frac{2}{p}})$ respectively.

\subsection{Effective central charge}

After all the simplifications mentioned above, the
$q$-series we shall consider is
\begin{equation}
\begin{split}
{\tilde{\chi}}_{1,1}^{(p)}(A,B;q) =\sum_{n=0}^{\infty}
\sum_{m_1,\dots,m_{p-3}=0}^{\infty}&\({-\frac{q^{\frac{1}{2}}}{A}}\)_n
\({-\frac{q^{\frac{1}{2}}}{B}}\)_n(AB)^n
\frac{q^{\vec{m}^T{C_{p-3}}\vec{m}-2nm_1+n^2}}{(q)_{2n}}\\
&\times\prod_{i=1}^{p-3}\Mult{(I_{p-3}\vec{m}
+n\vec{e}_1)_i}{2m_i}{q}.
\end{split}
\end{equation}
If the coefficients in this series 
${\tilde{\chi}}_{1,1}^{(p)}=\sum{a_M}q^M$ behave like $a_M\sim
{e^{2\pi\sqrt{M\tilde{c}/6}}}$ for large $M$, then
as $q\rightarrow1^{-}$, ${\tilde{\chi}}_{1,1}^{(p)}$ diverges like
${\tilde{q}}^{-\tilde{c}/24}$.
In other words one can obtain the `generalized' central charge $\tilde{c}$
from the asymptotic growth of the coefficient $a_M$.
The latter is computed by applying the saddle point method to
\begin{equation}
a_{M-1} = \oint\frac{dq}{2\pi{i}}{\tilde{\chi}}_{1,1}^{(p)}(A,B;q)q^{-M} =
\oint\frac{dq}{2\pi{i}}\sum_{n}\sum_{\vec{m}}
f(n,\vec{m};q).
\end{equation}
The saddle point occurs at the point where the derivatives of
\begin{multline}\label{e:f}
\log{f(n,\vec{m};q)} \approx 
\int_{0}^{n}\log\({1+\frac{q^k}{A}}\)dk+
\int_{0}^{n}\log\({1+\frac{q^k}{B}}\)dk+n\log(AB)\\
\shoveright{-\int_{0}^{2n}\log(1-q^k)dk
+(n^2-2nm_1+\vec{m}^T{C_{p-3}}\vec{m}-M)\log{q}}\\
+
\sum_{i=1}^{p-3}\({\int_0^{(I_{p-3}\vec{m}+n\vec{e}_1)_i}
-\int_0^{(I_{p-3}\vec{m}+n\vec{e}_1-2\vec{m})_i}-\int_0^{2m_i}}\)\log(1-q^k)dk
\end{multline}
with respect to $n$, $m_1, \dots, m_{p-3}$ and $q$ are all zero.
In deriving the expression in \eqref{e:f}, sums such as
$\log\{(q)_n\}$ and $\log\{({-q^{\frac{1}{2}}/A})_n\}$ were
approximated by integrals.
There are several ways to make this approximation. Ultimately,
the difference between the various approximation schemes is equivalent to
a difference in the linear terms in the exponent of $q$, and do not influence the quadratic
terms. Since the asymptotic growth is not expected to depend on the linear terms as
explained above, we have the freedom to use
the following two (different) approximations:
\begin{subequations}
\label{e:approx}
\begin{align}
\log\bigl\{(q)_n\bigr\}& \sim \int_0^{n}\log(1-q^k)dk
\quad\text{and}\\
\log\Biggl\{\({-\frac{q^{\frac{1}{2}}}{A}}\)_n\Biggr\}&
\sim \int_0^{n}\log\({1+\frac{q^k}{A}}\)dk.
\end{align}
\end{subequations}
This combination of approximations was chosen to simplify the algebra after
differentiation.

Let us define
\begin{equation}
v_i = q^{-2m_i} \quad \text{and} \quad w_i = q^{(I_{p-3}\vec{m}+n\vec{e}_1)_i}.
\end{equation}
The differentiation with respect to $m_i$ and $n$ produced 
the following set of relations for their
saddle point values $\b{m}_i$ and $\b{n}$:
\begin{subequations}
\label{e:cond1}
\begin{align}
(1-y_i)^2& = \prod_{j=1}^{p-3}\,y_j^{I_{ij}},\\
q^{2\b{n}\delta_{1,i}}(1-x_i)^2&
= \prod_{j=1}^{p-3}\,x_j^{I_{ij}},\\
(1-q^{2\b{n}})^2& = (A+q^{\b{n}})(B+q^{\b{n}})q^{2\b{n}}x_1,
\end{align}
\end{subequations}
where
\begin{equation}
x_i = \frac{(1-\b{w}_i)\b{v}_i}{1-\b{v}_i\b{w}_i}
\quad\text{and}\quad
y_i = \frac{(1-\b{w}_i)}{1-\b{v}_i\b{w}_i}.
\end{equation}
It is easy to show that
in the special cases of \eqref{e:case2}, \eqref{e:cond1}
reduces to a system of algebraic equations governed by the algebras
$A_{p-2}$, $A_{p-1}$ and $D_{p-1}$ respectively. For these algebras,
the corresponding
systems of equations are solved in the literature, and are known to
be related to the Thermodynamic Bethe Ansatz (TBA) approach (see for example
\cite{35,36,37,38,39}).
Here we can easily write down the solution for $y_i$ as
\begin{equation}
y_i = \frac{{\sin}^2(1+i)\frac{\pi}{p}}{{\sin}^2\frac{\pi}{p}}.
\end{equation}
One can also show that
\begin{equation}
x_i = \frac{\sin^2(p-1-i)\h}
{\sin^2\h}
\end{equation}
satisfies \eqref{e:cond1} with the closure conditions
\begin{subequations}
\begin{align}
x_{p-2}& = 1,\\
x_0& =\frac{\sin^2(p-1)\h}{\sin^2\h}\quad=q^{-2\b{n}},\\
x_{-1}& =\frac{\sin^2p\h}{\sin^2\h}\quad=(1+Aq^{-\b{n}})(1+Bq^{-\b{n}}).\\
\end{align}
\end{subequations}
The parameter $\h$ is related to $A$ and $B$ by the relation
\begin{equation}
\label{e:relation}
(A+B)\sin\h + AB\sin(p-1)\h = \sin(p+1)\h.
\end{equation}

To compute $\log{f(n,\vec{m};q)}$
at the stationary point with respect to $m_i$ and $n$, we first
rewrite it using the relations
\begin{subequations}
\label{e:simple}
\begin{align}
\int_{0}^{\b{z}}\log{(1-q^k)}dk& =
\frac{1}{\log{q}}\Bigl[L(1-q^{\b{z}})+
\frac{1}{2}\log{(1-q^{\b{z}})}\log{q^{\b{z}}}\Bigr];\\
\int_{0}^{\b{z}}\log{\(1+\frac{q^k}{A}\)}dk& =
\frac{1}{\log{q}}\Bigl[L\({\frac{q^{\b{z}}}{q^{\b{z}}+A}}\)
-L\({\frac{1}{1+A}}\) \notag \\
&\quad +\frac{1}{2}\log{A}\log{\({\frac{1+A}
{q^{\b{z}}+A}}\)}\Bigr]
+\frac{\b{z}}{2}\log{\({1+\frac{q^{\b{z}}}{A}}\)}.
\end{align}
\end{subequations}
The Rogers dilogarithm in \eqref{e:simple} is defined by \cite{Lewin}
\begin{equation}
L(z) = Li_2(z) + \frac{1}{2}\log{z}\log{(1-z)};\quad
Li_2(z) = -\int_{0}^{z}\frac{\log(1-w)}{w}dw
\end{equation}
and $L(1) = \frac{{\pi}^2}{6}$.
The five terms relation for the dilogarithm in our case can be written as
\begin{equation}
L(1-w_i)-L(1-v_iw_i)-L(1-v_i^{-1}) =
L(1-y_i^{-1})-L(1-x_i^{-1}).
\end{equation}
Hence we have 
\begin{equation}
\label{e:logf}
\log{f(n,\vec{m};q)}\Bigl\rvert_{\substack{\vec{m}=\vec{\b{m}}\\ n=\b{n}}}
\,\approx\, -M\log{q} - \frac{{\pi}^2\tilde{c}(A,B|p)}{6\log{q}},
\end{equation}
where
\begin{equation}
\label{e:c}
\begin{split}
\tilde{c}(A,B|p) = &\frac{1}{L(1)}\(L\(\frac{1}{1+A}\)+L\(\frac{1}{1+B}\)
+L\(1-\frac{1}{x_0}\)
-L\(\frac{1}{1+\sqrt{x_0}A}\)\\
&-L\(\frac{1}{1+\sqrt{x_0}B}\)
+\sum_{i=1}^{p-3}\Bigl[L\(1-\frac{1}{x_i}\)-L\(1-\frac{1}{y_i}\)\Bigr]\\
&+\frac{1}{2}\log{A}\log\(\frac{1+\sqrt{x_0}A}{1+A}\)
+\frac{1}{2}\log{B}\log\(\frac{1+\sqrt{x_0}B}{1+B}\)
\).
\end{split}
\end{equation}

By differentiating \eqref{e:logf} with respect to $q$, we found the saddle point
value of $q$ to be
\begin{equation}
\b{q} = e^{-\sqrt{{\pi}^2\tilde{c}/6M}}.
\end{equation}
This leads to the expected asymptotic behavior of $a_M$ for large $M$, and
hence we can interpret $\tilde{c}(A,B|p)$ as a `generalized' effective central
charge for \eqref{e:genchar}. The sums in \eqref{e:c} can be further
simplified using dilogarithm sum rules \cite{Lewin,41} to yield
\begin{equation}
\sum_{i=1}^{p-3}L\(1-\frac{1}{y_{i}}\) =
(p-5+\frac{6}{p})L(1); \quad \text{and}
\end{equation}
\begin{multline}
L\(1-\frac{1}{x_0}\)+\sum_{i=1}^{p-3}\Bigl[L\(1-\frac{1}{x_i}\)\Bigr]
= (p-1)L(1) - p(p-1){\h}^2\\
+2Li_2\(-\frac{\sin(p-1)\h}{\sin\h},p\h\) +
\log\(\frac{\sin(p-1)\h}{\sin\h}\)\log\(\frac{\sin{p}\h}{\sin\h}\),
\end{multline}
where
\begin{equation}
Li_2(r,\h) = Re\{Li_2(re^{i\h})\} = -\frac{1}{2}\int_0^{r}
\frac{\log(1-2x\cos\h+x^2)}{x}dx.
\end{equation}
The resultant expression for the `generalized' central charge is
\begin{equation}
\label{e:effc}
\begin{split}
\tilde{c}(A,B|p) =& \frac{1}{L(1)}\Bigl[Li_2(-A)+Li_2(-B)
-Li_2\(-A\frac{\sin(p-1)\h}{\sin\h}\)\\
&-Li_2\(-B\frac{\sin(p-1)\h}{\sin\h}\)
+(4-\frac{6}{p})L(1) - p(p-1)\h^2\\
&+ 2Li_2\(-\frac{\sin(p-1)\h}{\sin\h},p\h\)\Bigr].
\end{split}
\end{equation}
The simple expression in \eqref{e:effc} is the main result of this
letter. It gives the effective central charge associated with the
`generalized' character formula \eqref{e:genchar} in terms of
dilogarithm functions. The expressions in \eqref{e:c} and
\eqref{e:effc} are valid for $A\ge0$ and $B\ge0$.

\subsection{Special cases}

Consider the domain
$A=0$, and $B=\frac{\sin(p+1)\h}{\sin\h}$ follows from \eqref{e:relation}.
To implement the special case (I), we take the limit
$B\rightarrow0$, thus yielding $\h=\frac{\pi}{p+1}$.
Consequently by using the identity \cite{Lewin}
\begin{equation*}
Li_2(2\cos{\h},\h) = (\frac{\pi}{2}-\h)^2,
\end{equation*}
we obtained
$\tilde{c}(0,0|p)=1-\frac{6}{p(p+1)}$, which is the central
charge of the unitary model $M(p,p+1)$.
In the limit (II), taking $B=1$ we found $\h=\frac{\pi}{p+2}$.
Using the limit
\begin{equation*}
Li_2\({-\frac{\sin{(p-1)\h}}{\sin{\h}}}\)
+2Li_2\({-\frac{\sin{(p-1)\h}}{\sin{\h}},p\h}\)\Biggl\rvert_{\h=\frac{\pi}{p+2}}
= \frac{2}{3}{\pi}^2-2\frac{(2p+1)}{(p+2)^2}{\pi}^2,
\end{equation*}
we recovers
the central charge
of the $N=1$ unitary model
$\tilde{c}(0,1|p)=\frac{3}{2}-\frac{12}{p(p+2)}$.

It is interesting
that $\tilde{c}(0,B|p)$ is a smooth monotonic function of $B$ between
the above two limits.
In particular for the case of $p=3$, we have a function which
connects the central charges of the Ising and Tricritical Ising model%
\footnote{In this case, the `generalized' character \eqref{e:genchar}
take us from the Tricritical Ising character $\chi_{1,1}+\chi_{1,4}$
to the Ising character $\chi_{1,1}$.}%
. Hence it is desirable to compare $\tilde{c}(0,B|3)$ with the known
ground state scaling function $\mathcal{C}(r)$ obtained from TBA \cite{6}.
Recall that the latter is 
a function of a scaling parameter $r$, with UV limit 
$(r\rightarrow0)$ $\frac{7}{10}$
and IR limit $(r\rightarrow\infty)$
$\frac{1}{2}$ respectively.
Therefore to compare the two expressions, we need to find a parameterization
of the
variable $B$ in terms of $r$.
This can always be done since one can in principle
invert the function $\tilde{c}(0,B|3)$ to obtain the parameterization
$B(r)={\tilde{c}}^{-1}(\mathcal{C}(r))$.
However we were unable to express
$B(r)$ in a simple and closed form. This is perhaps not surprising since
$\mathcal{C}(r)$ is written as an integral involving two pseudo-energies
$\varepsilon_1$ and $\varepsilon_2$,
who in turn are given by two coupled integral equations involving $r$.
Only in the UV or IR limits do we get a simplification of the integral
equations, which then allow us to write $\mathcal{C}$ in terms of dilogarithms
\cite{6}.
Hence the parameterization $B(r)$,
which yields an expression for $\mathcal{C}(r)$ in terms of
dilogarithms for general $r$, is likely to be complicated.
It is also unclear at this stage whether this parameterization admits any physical
interpretations.

Now let us focus on the other domain $A=B$.
The relation \eqref{e:relation} tells us
$A=\frac{\cos(p+1)\h/2}{\cos(p-1)\h/2}$.
Of course in the limiting case (I), $A\rightarrow0$,
we found $\h=\frac{\pi}{p+1}$
as before.
Once again $\tilde{c}(A,A|p)$ is a smooth monotonic function of $A$.
For the $N=2$ supersymmetric limit (III), taking $A\rightarrow1$, we get
$\h=0$ and $\tilde{c}(1,1|p)=3(1-\frac{2}{p})$ as expected.
Hence $\tilde{c}(A,B|p)$ indeed give us a function which interpolates between
the central charges of an unitary model and its supersymmetric counterparts.

\section{Discussion}

In this work, we have studied the asymptotic behavior of the `generalized'
character formula $\chi_{s,r}^{(p)}(\ra,\r2;q)$ \eqref{e:genchar} in the
limit $q\rightarrow1^{-}$. In this limit, we show that
the $q$-series diverges like
$\tilde{q}^{-\frac{\tilde{c}}{24}}$ and we found a simple expression
\eqref{e:effc}
for the `generalized' effective central charge $\tilde{c}$ in terms of
dilogarithms. In the limiting cases (I), (II) and (III), $\tilde{c}$
yields the central charges of the unitary models and their
supersymmetric counterparts.

Having stated our conclusion, we shall take the liberty to indulge in
a some (pure) speculations. Of course it is not surprising that we can find a
function which reproduces the correct central charges in the various
limits. Indeed $\chi_{s,r}^{(p)}(\ra,\r2;q)$ also becomes the corresponding
CFT
characters in these limits. But what is {\em a priori} not expected from the 
Andrews-Bailey construction is that the `generalized' character
\eqref{e:genchar} would
exhibit the asymptotic behavior found in \S 3. The prediction
for this behavior was based on the assumption that \eqref{e:genchar}
gives the partition function for some quantum field
theory. This field theory
must be invariant under interchanging the roles of space and
time.
This suggests that $\chi_{s,r}^{(p)}(\ra,\r2;q)$,
when multiplied by a suitable factor $q^{\mathcal{D}_{s,r}(\ra,\r2|p)}$,
may be modular
covariant. It would be very interesting to show directly from
\eqref{e:genchar} that
\begin{equation*}
{q}^{\mathcal{D}_{s,r}(\ra,\r2|p)}
\chi_{s,r}^{(p)}(\ra,\r2;q) = \sum_{s^{\prime},r^{\prime}}
S_{s,r}^{s^{\prime},r^{\prime}}(\ra,\r2|p)
{\tilde{q}}^{\mathcal{D}_{s^{\prime},r^{\prime}}(\ra,\r2|p)}
\chi_{s^{\prime},r^{\prime}}^{(p)}(\ra,\r2;\tilde{q})
\end{equation*}
for some `generalized' $S$-matrix.
Presumably the elements of this
$S$-matrix (if it exits) can be calculated from the non-perturbative
corrections to the saddle point \cite{tel,nahm}. This computation would
be much more involved than that in \S 3 since the elements of $S$
depend on $r$ and $s$.

Another interesting puzzle is the nature of the quantum field theory with
the Hamiltonian $\mathcal{H_P}$ discussed in \S 3. From \eqref{e:Ec}, the
ground state energy of $\mathcal{H_P}(\ra,\r2|p)$
is proportional to $\tilde{c}(A,B|p)$ which is written in terms of
dilogarithms.
This seems to indicate that $\mathcal{H_P}(\ra,\r2|p)$ is
the Hamiltonian for a (maybe irrational) CFT which interpolates between
the $N=0$, $N=1$ and $N=2$ unitary models. It is known in some cases 
that by varying the closed boundary condition $P$ around the cylinder,
one can interpolate continuously between several CFTs \cite{zuber}.
Examples include the $Q$-Potts and $O(n)$ models \cite{col},
and the Ising model with defect lines (see for example \cite{defect}).
Whether one can obtain the `generalized' character \eqref{e:genchar}
by varying the cylindrical boundary conditions of the
$N=2$ unitary model is worth investigating.  

\subsection*{Acknowledgements}
The author is grateful to O. Foda and O. Warnaar for
reading the manuscript and many valuable discussions.
I would also like to thank R. Kedem, A. Kirillov, C. Pisani
and D. Pihan for useful conversations. This work is supported
by the Australian Research Council.

\bibliographystyle{amsplain}

\end{document}